\documentclass[12pt,preprint]{aastex}

\shorttitle{GRBs and AGASA anisotropy at $10^{18}$ eV}
\shortauthors{Biermann et al.}

\begin{document}
%%%%%% text sizes etc %%%%%%%%%%%%%%%%%%%%%
%\textheight23cm \textwidth15.5cm \oddsidemargin0.5cm
%\evensidemargin0cm
%\parskip1ex
%\pagestyle{myheadings}
%%%%%%%%%%%%%%%%%%%%%%%%%%%%%%%%%%%%%%%%%%%

\newcommand{\D}{\displaystyle} %normal formulas
\newcommand{\T}{\textstyle} %for large font
\newcommand{\SC}{\scriptstyle} %footnote
\newcommand{\SSC}{\scriptscriptstyle} %footnote to footnote

\def\AJ{{\it Astron. J.} }
\def\ARAA{{\it Annual Rev. of Astron. \& Astrophys.} }
\def\ApJ{{\it Astrophys. J.} }
\def\ApJL{{\it Astrophys. J. Letters} }
\def\ApJS{{\it Astrophys. J. Suppl.} }
\def\ApP{{\it Astropart. Phys.} }
\def\AA{{\it Astron. \& Astroph.} }
\def\AAR{{\it Astron. \& Astroph. Rev.} }
\def\AAL{{\it Astron. \& Astroph. Letters} }
\def\JGR{{\it Journ. of Geophys. Res.}}
\def\JHEP{{\it Journal of High Energy Physics} }
\def\JPhG{{\it Journ. of Physics} {\bf G} }
\def\PhFl{{\it Phys. of Fluids} }
\def\PR{{\it Phys. Rev.} }
\def\PRD{{\it Phys. Rev.} {\bf D} }
\def\PRL{{\it Phys. Rev. Letters} }
\def\Nature{{\it Nature} }
\def\MNRAS{{\it Month. Not. Roy. Astr. Soc.} }
\def\ZA{{\it Zeitschr. f{\"u}r Astrophys.} }
\def\ZFN{{\it Zeitschr. f{\"u}r Naturforsch.} }
\def\etal{{\it et al.}}

\hyphenation{mono-chro-matic  sour-ces  Wein-berg chang-es
Strah-lung dis-tri-bu-tion com-po-si-tion elec-tro-mag-ne-tic
ex-tra-galactic ap-prox-i-ma-tion nu-cle-o-syn-the-sis
re-spec-tive-ly su-per-nova su-per-novae su-per-nova-shocks
con-vec-tive down-wards es-ti-ma-ted frag-ments
grav-i-ta-tion-al-ly el-e-ments me-di-um ob-ser-va-tions
tur-bul-ence sec-ond-ary in-ter-action in-ter-stellar spall-ation
ar-gu-ment de-pen-dence sig-nif-i-cant-ly in-flu-enc-ed par-ti-cle
sim-plic-i-ty nu-cle-ar smash-es iso-topes in-ject-ed
in-di-vid-u-al nor-mal-iza-tion lon-ger con-stant sta-tion-ary
sta-tion-ar-i-ty spec-trum pro-por-tion-al cos-mic re-turn
ob-ser-va-tion-al es-ti-mate switch-over grav-i-ta-tion-al
super-galactic com-po-nent com-po-nents prob-a-bly
cos-mo-log-ical-ly Kron-berg Rech-nun-gen La-dungs-trenn-ung
ins-be-son-dere Mag-net-fel-der bro-deln-de}
%Formeln
\def\simle{\lower 2pt \hbox {$\buildrel < \over {\scriptstyle \sim }$}}
\def\simge{\lower 2pt \hbox {$\buildrel > \over {\scriptstyle \sim }$}}

\title{The last Gamma Ray Burst in our Galaxy?\\
On the observed cosmic ray excess at particle energy $10^{18}$ eV}

\author{Peter L. Biermann\footnote{Max-Planck Inst. for Radioastronomy
and Dept. Phys. and Astron., Univ. of Bonn, Germany,
plbiermann@mpifr-bonn.mpg.de} }

\author{Gustavo Medina Tanco\footnote{Inst. Astron. e Geof., Univ.
de Sao Paulo, Brasil, gustavo@astro.iag.usp.br} }

\author{Ralph Engel\footnote{Forschungszentrum Karlsruhe, Karlsruhe,
Germany, Ralph.Engel@ik.fzk.de} }

\author{Giovanna Pugliese\footnote{European Southern Obs., Munich,
Germany, gpuglies@eso.org} }

%
%  version of May 7, 2003  17h49 - now after changes by RE, further
%  modifications by PLB May 12, 2003 15h40
%  after referee report of July 24, further corrections by PLB and RE
%  here version of July 27, 10h24
%

\begin{abstract}
Here we propose that the excess flux of particle events of energy
near $10^{18}$ eV from the direction of the Galactic Center region
is due to the production of cosmic rays by the last few Gamma Ray
Bursts in our Galaxy.  The basic idea is that protons get
accelerated inside Gamma Ray Bursts, then get ejected as neutrons,
decay and so turn back into protons, meander around the inner
Galaxy for some time, and then interact again, turning back to
neutrons to be observed at our distance from the Galactic Center
region, where most star formation is happening in our Galaxy.  We
demonstrate that this suggestion leads to a successful
interpretation of the data, within the uncertainties of cosmic ray
transport time scales in the inner Galaxy, and in conjunction with
many arguments in the literature.
\end{abstract}


\section{Introduction}

For some time now, the detection of an excess in $10^{18}$ eV
cosmic rays from the general direction of the Galactic Center (GC)
by the air shower array AGASA \cite{AGASA99GC} has been a special
riddle in Galactic cosmic ray research.  The angular region of the
excess observed by AGASA and SUGAR is illustrated in Fig.~1.

%%%%%%%%%%%%%%%to insert figure 1 %%%%%%%%
\begin{figure}
\centering\rotatebox{0}{\resizebox{13cm}{!}%
{\includegraphics{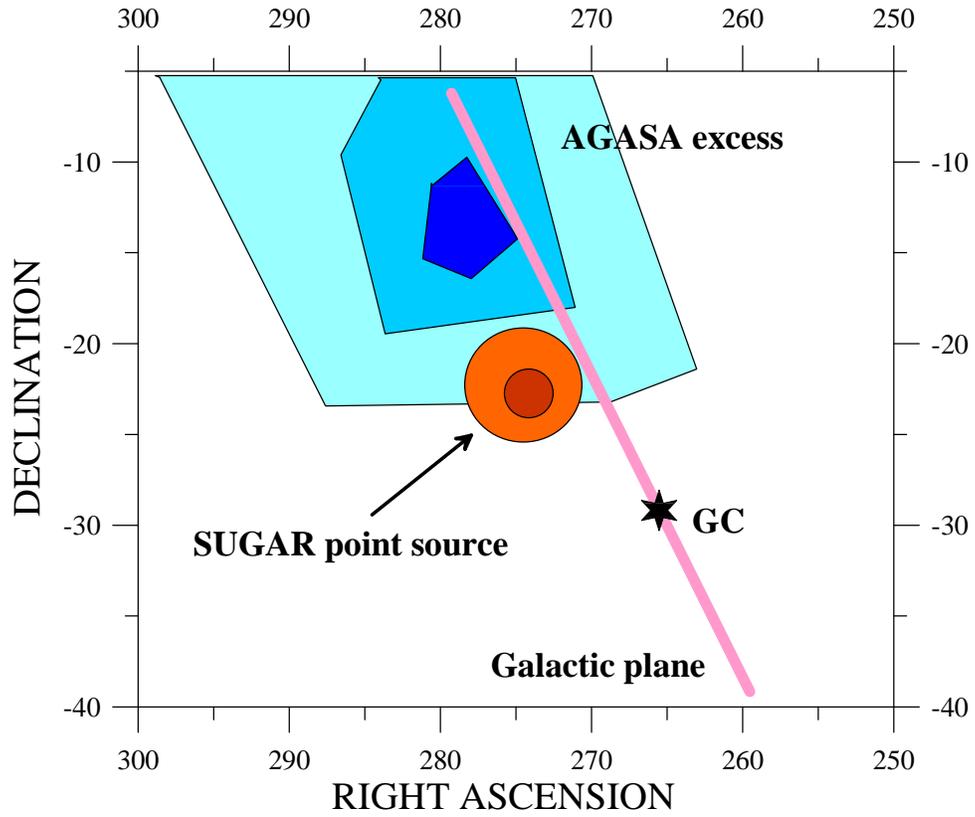}}} \caption{Cartoon map of AGASA and
SUGAR data. Indicated are the regions from which an excess of
cosmic rays is observed. In addition the location of the Galactic
Center (GC) and the Galactic plane is marked. Right ascension and
declination are both given in degrees.}
\end{figure}
%%%%%%%%%%%%%%%%%%%%%%%%%%%%%%%%%%%%%%%%%

Gamma Ray Bursts (GRB) have long been argued to produce high
energy cosmic rays. In this paper we will show what the
observational signature should be and try to demonstrate that the
AGASA excess can be attributed to the last one, or last few GRB in
our inner Galaxy.

%\subsection{The data}

%First we summarize the features of the cosmic ray excess known so
%far.

There is a significant excess around $10^{18}$ eV, seen by two
different experiments, AGASA and SUGAR, of cosmic rays coming from
the general direction of the GC region, that cannot be attributed
to the expected gradient of cosmic ray sources and distribution.
This excess has been supported at the time by the air fluorescence
detector Fly's Eye \cite{FE99}.  However the sky coverage of AGASA
does not include the GC, while that of SUGAR does \cite{Clay01}.
The SUGAR data suggest a point source to within their spatial
resolution, while AGASA shows an extended source.

The excess starts to be significant around $3 \times 10^{17}$ eV,
peaks near $10^{18}$ eV, and cuts off sharply at about $3 \times
10^{18}$ eV, see \cite{Teshima01a}.

%%%%%%%%%%%%%%%to insert figure 2 %%%%%%%%
%\begin{figure}
%\centering\rotatebox{0}{\resizebox{8 cm}{!}%
%{\includegraphics{f2.c.eps}}} \caption{In this graph
%\cite{Teshima01a} we show the significance of the first harmonic
%analysis in right ascension as a function of threshold energy;
%what is plotted is the negative logarithm of the probability that
%this flux is random. In the ordinate the chance probability of the
%observed excess is shown. In particular, we can see that the
%significance of the excess is very large in a narrow region around
%$10^{18}$ eV. The chance probability of the observed arrival
%directions  being compatible with the expected isotropic
%distribution is $\sim 10^{-5}$ at this energy.}
%\end{figure}
%%%%%%%%%%%%%%%%%%%%%%%%%%%%%%%%%%%%%%%%%

The flux of the excess particles can be turned into a luminosity
of particles beyond $10^{18}$ eV of about $4 \times 10^{30}$
erg/s.  Since AGASA cannot observe the entire region, this
inferred luminosity must be a lower limit, with the true
luminosity possibly being a factor of 3 - 10 larger.

%\subsection{Immediate conclusions}

%There are a number of immediate consequences and implications:

As suggested by the AGASA Collab. these events may arise from
neutrons  \cite{AGASA99GC}; the possibility that they are due to
photons was discussed and discarded by \cite{Bellido02} on the
basis of SUGAR data. Firstly, neutrons are not deflected due to
the Galactic magnetic fields. Secondly, a peak at exactly
$10^{18}$ eV corresponds to the distance to the GC region (in
distance from here between 5 and 11 kpc, using the region of
highest star formation rate of about 3 kpc around the GC
\cite{Gusten83}), folding the neutron decay with an injected power
law spectrum. Only with neutrons would there be a lower limit in
energy, above which there can be significant flux. With photons
there is no such threshold.

In this case the original proton must have had an energy about
three times that observed, and so we require proton energies of at
least up to $6 \times 10^{18}$ eV.

There are three main mechanisms and respective sites to accelerate
particles in the Galaxy: supernova explosions either in the
interstellar medium, in young and hot star bubbles, or in massive
star winds.  In any of the three cases such an energy per nucleon
cannot be reached for any reasonable parameter of shock velocity
and/or magnetic field
\cite{Lagage+C83,Jokipii87,CRII,CRIII,CalgaryCR}.

%\subsection{The Gamma Ray Burst hypothesis}

The only way to accelerate particles to such an energy per nucleon
in a normal galaxy as ours is relativistic shocks. Such
relativistic shocks are produced in GRBs \cite{Vietri98,Piran99}.
GRBs are believed to occur in every galaxy in some small fraction
of all supernovae, of order $10^{-4}$, arising from the final
evolutionary stages of very massive stars, either in a binary
system or as a single star. Therefore we propose to ask: What are
GRBs expected to produce in terms of energetic particles?

As shown in \cite{Rachen98a}, because of adiabatic losses, the
highest energy particles that emerge from a GRB are mostly
neutrons; protons are captive in the magnetic field and so suffer
extensive adiabatic losses on the way out. These neutrons will
decay after some distance, turning into protons, which are then
caught by the magnetic field in the Galaxy, and rumble around with
a rather short residence time scale.  There is a small, but finite
probability that they will produce a neutron again in interactions
with the interstellar medium.  These secondary neutrons then could
travel undeflected to us to be observed.

We will try to follow the neutrons originally ejected from a GRB.
Fig.~2 shows our concept in three key time steps.

%%%%%%%%%%%%%%%to insert figure 2 %%%%%%%%
\begin{figure}
\centering\rotatebox{0}{\resizebox{13.0 cm}{!}%
{\includegraphics{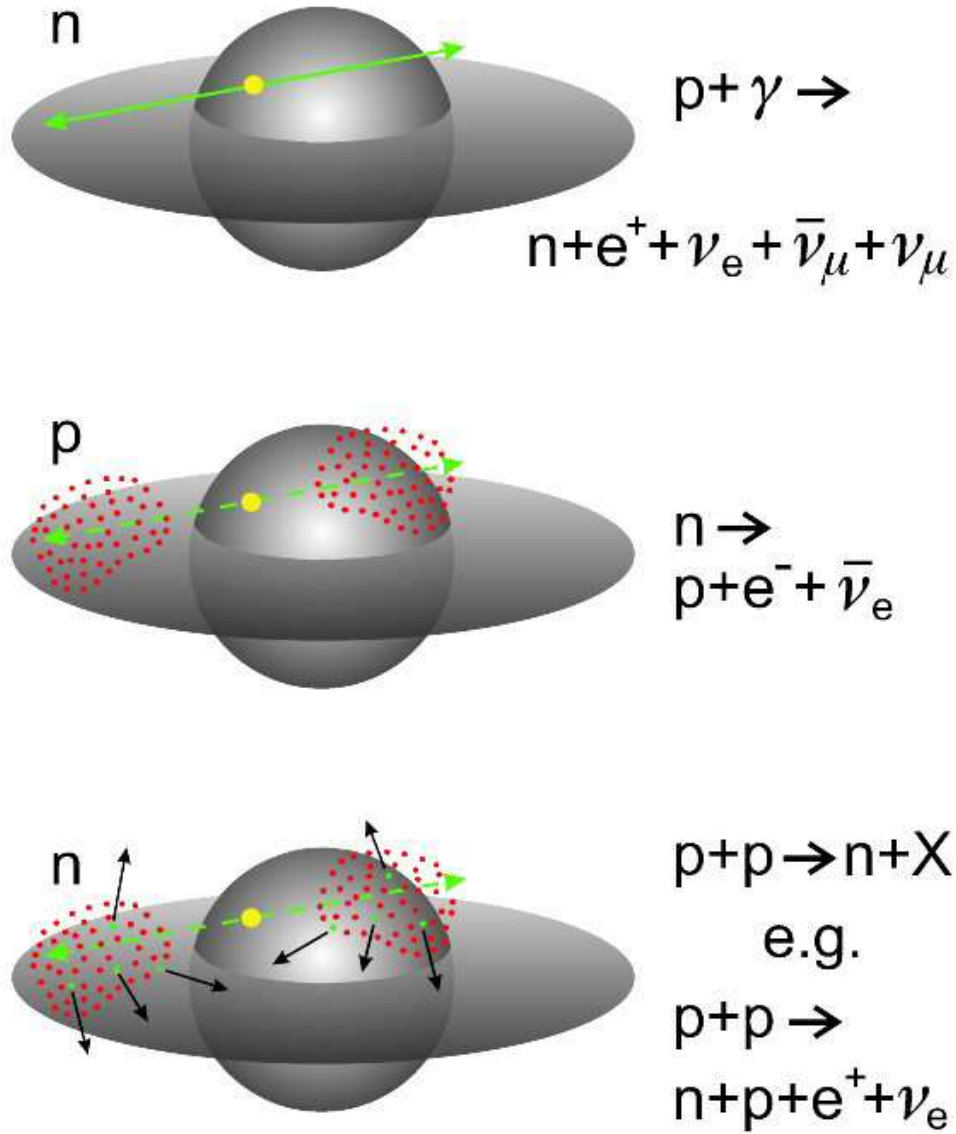}}} \caption{In this graph we
illustrate the scenario which we propose in three key time steps.
First, neutron ejection, second, neutron decay, and finally,
proton interaction leading to new neutrons being observable from
Earth. The GRB is taken to happen in the thin disk within three
kpc from the GC, seen here in projection together with the old
stellar bulge.}
\end{figure}
%%%%%%%%%%%%%%%%%%%%%%%%%%%%%%%%%%%%%%%%%

\section{The last GRB event in our Galaxy}

In order to estimate the remaining traces of any activity of
cosmic rays ejected and/or produced by GRBs, we discuss below
various probabilities, the residence time for the protons
resulting from the decay of the neutrons ejected, and present
spectral and spatial arguments.

We will somewhat arbitrarily adopt here a GRB rate of 1 per
$10^{6}$ years for the inner Galaxy  (considering \cite{Piran01},
this may seem pessimistic, but considering \cite{Frail01}, even
optimistic).

The residence time scale for energetic protons in the Solar
neighborhood is around, \cite{Gaisser90,Fe94}, $ \tau_{res} \, =
\, 2 \times 10^{7} \, {\rm yrs} \, ({E}/{{10^{9} \, \rm
eV}})^{-1/3} $ and so is about $2 \times 10^{4}$ yrs at $10^{18}$
eV .   Obviously, near the GC this time scale could be different.
Since the magnetic field strength is higher, and the path to the
outer parts of the Galactic halo is also larger it is likely that
the time is longer, and so we adopt $10^{5}$ years. After an
injection event the number of cosmic rays at some energy will
decay in a 3D-slab diffusion approximation with $t^{-3/2}$ for
times up to the diffusive reservoir time, and with $t^{-5/2}$
thereafter.  For the nominal injection time scale of every
$10^{6}$ yrs, the expected diminution factor is about  1/300 at
again the typical time to the next GRB. The decay length of a
neutron at $10^{18}$ eV exceeds the thickness of the GC thin
disk-like region of high density and high star formation rate.
Using  a 3 kpc radius of the region of  high interstellar medium
density and high star formation rate in the Galactic disk about
the GC and a decay length of 10 kpc this leads to a suppression
factor of about 1/3. Another important aspect of the model is the
probability of a proton to produce again an energetic secondary
neutron in a collision with a nucleus of the interstellar medium
in the inner Galaxy.  Using conservatively the typical grammage
inferred from the boron to carbon ratio in cosmic rays,  with the
Solar neighborhood again as a reference, this probability is of
the order of a few percent, estimated to 1/20 (simulated with
cross sections from \cite{PDG02}). Next we compare in the
following the flux of particles above $10^{18}$ eV with the flux
at all cosmic ray energies combined. Assuming the injection
spectrum to be around $E^{-2.2}$ at first produces a diminution of
the flux beyond $10^{18}$ eV by a factor of about 1/100
\cite{Rachen98a,BeOs98}. Photons are produced in decays of neutral
pions,  but the typical energy of secondary pions is lower than
that of energetic secondary protons and neutrons. Thus the final
photon energy is lower and so may become submerged in the higher
cosmic ray flux at lower energies.  In Fig.~3 we show a simulation
for a power law injection spectrum with an arbitrary exponential
cutoff in energy at $E_{\rm cut} = 10^{18}$ eV:  the differential
flux is $E^{-2.2} \; {\rm for} \; E < E_{\rm cut}$ and $E^{-2.2}
\exp\left[-(E-E_{\rm cut})/E_{\rm cut}\right]   \; {\rm for} \; E
\ge E_{\rm cut} $. Down to an energy of about a factor ten below
the cutoff energy the effect of the cutoff is clearly  visible in
the ratio of the various fluxes, and below that energy, the ratio
approaches a constant. What is the probability that a GRB jet
points more or less along the Galactic plane? As seen from the GC
region the Galactic plane is quite thick, and so maybe half the
sky has an appreciable column density -- see, e.g., \cite{ZMW90}
-- leading to a factor of 1/2. And, finally, what is the total CR
production by a ``typical'' GRB?  We adopt here the generous
number of $10^{51}$ erg in pure cosmic rays from
\cite{Piran99,GRB-2,VB02}.

%%%%%%%%%%%%%%%to insert figure 3 %%%%%%%%
\begin{figure}
\centering\rotatebox{0}{\resizebox{13 cm}{!}%
{\includegraphics{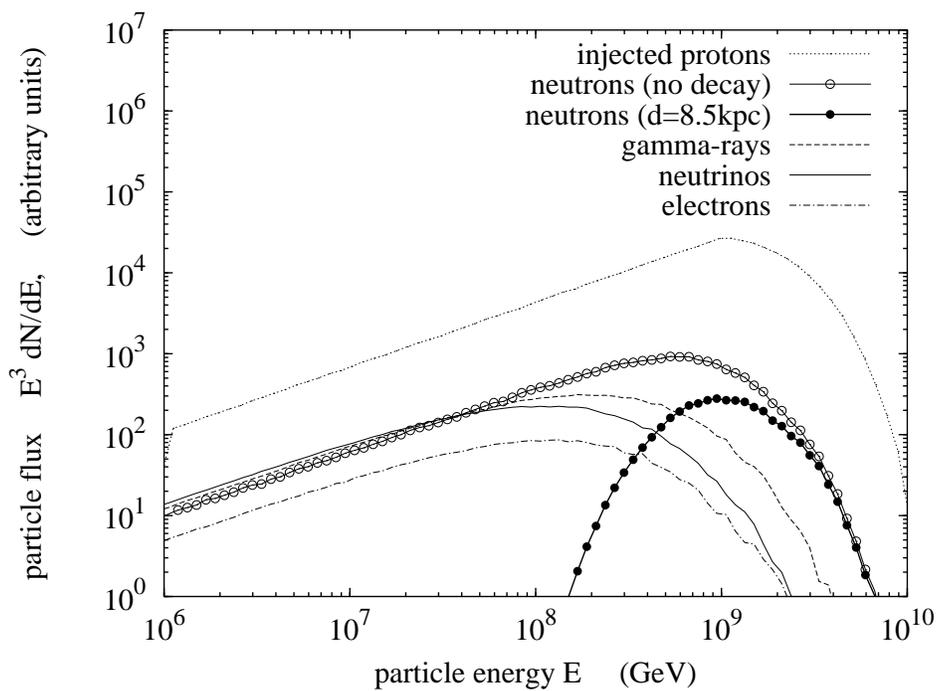}}} \caption{Resulting spectra of
secondary particles, such as secondary neutrons/antineutrons,
gamma-rays, electrons/positrons and neutrinos. For comparison the
secondary neutron spectrum near the Galactic Center (no decay) and
the local one for a distance of $d=8.5$ kpc from the production
site is shown.}
\end{figure}
%%%%%%%%%%%%%%%%%%%%%%%%%%%%%%%%%%%%%%%%%

Therefore the expected flux today, from the last GRB occurring
$10^{6}$ yrs ago is now $10^{31}$ erg/s.  This is  just above what
is observed, and so allowing for uncertainties mainly  due to the
limited sky coverage of AGASA, a very plausible estimate to
explain the data.

The observed spectrum would be completely dominated by the two
step propagation of the secondary neutrons in such a picture.
Therefore the spectrum is the folding of the production spectrum,
with the decay probability inside the available space, so a hump
from the minimum distance to get any neutrons, see Fig.3, to the
maximum energy possible from GRB productions.

Last we wish to ask whether it could be that we observe the effect
of a few GRBs.   The most important parameters are 1) the time
since the GRB event, and 2) the orientation with respect to the
Galactic plane.  So adopting a picket-fence model for the timing
of subsequent GRBs, we note that the flux from a GRB that was
twice as old is decreased by a factor of about 5 with respect to
the younger one, and a GRB at three times the age shows a flux
that is a factor of 16 down.

We conclude that the observed distribution is rather likely to be
the result of several GRB events in the GC region.

\section{Predictions and tests}

Large numbers of photons, electrons and neutrinos are produced in
the collisions that give rise to the second generation neutrons in
such a picture.  However, their mean energy is small relative to
that of the neutrons. The fluxes of photons, electrons and
neutrinos are expected to follow an energy spectrum similar to
that of the interaction protons. Furthermore, photons and
neutrinos map out the spatial and angular distribution of the
proton interactions. It is convenient to express the secondary
particle spectra in terms of the primary proton spectrum by
multiplying it with  appropriate reduction factors.  Simulations
using the Monte Carlo event generator SIBYLL 2.1
\cite{Sybill94,Engel99a} predict the following reduction factors:
for secondary protons and antiprotons 0.27, for neutrons and
antineutrons 0.09, for photons 0.11, for electron-positron pairs
0.05, and for neutrinos (all flavors) 0.13. All these numbers are
normalized to a primary proton spectrum, using a power law of
$E^{-2.2}$, and the energy range $10^{17}$ - $10^{18}$ eV. These
numbers are the ratio of the fluxes far below the upper energy
cutoff. Observable is the ratio of the uncharged components, e.g.,
the ratio of neutrons to gamma-rays, which is here close to 1;
however, as Fig. 3 shows, near to the upper cutoff the photons
drop off earlier than the neutrons.  The curves can be shifted in
energy for any other assumption of maximum energy, since they will
look the same relative to maximum energy (here $10^{18}$ eV with a
following exponential cutoff).

%%%%%%%%%%%%%%%to insert figure 4 %%%%%%%%
\begin{figure}
\centering\rotatebox{0}{\resizebox{13 cm}{!}%
{\includegraphics{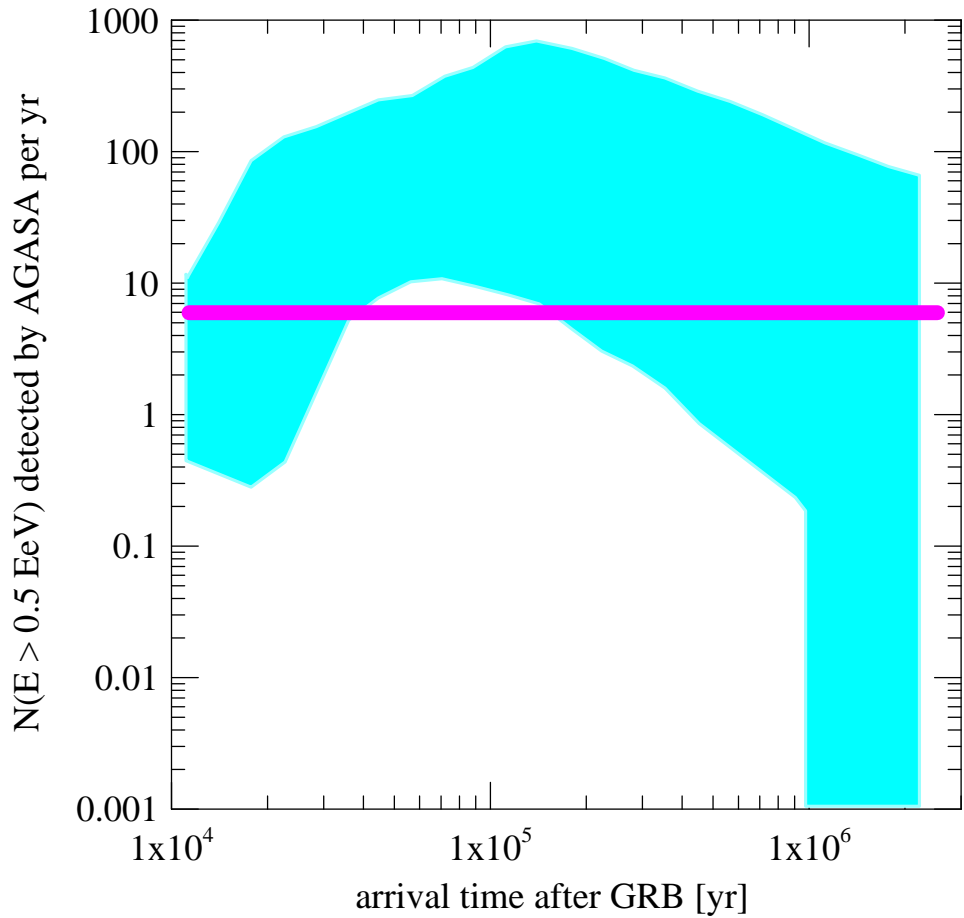}}} \caption{Number of neutrons due to
GBRs that would be detected in excess to an isotropic background
distribution. The simulation is based on 1000 random realizations
of GRBs with uniformly distributed jet orientation in the inner
Galaxy.  The shaded band shows the 95 \% CL region for the flux of
particles above  $5 \times 10^{17}$ eV as a function of time
detected by an AGASA-like instrument (same aperture and same
declination dependence of the exposure). The line indicates the
excess actually observed by AGASA.}
\end{figure}
%%%%%%%%%%%%%%%%%%%%%%%%%%%%%%%%%%%%%%%%%

To see an appreciable flux of neutrons peaking near $10^{18}$ eV
with a visible extension to $2 \times 10^{18}$ eV requires that
the primary proton/neutron flux extends to at least about $6
\times 10^{18}$ eV.  A measurement of the ratio of neutrons to
photons, with a simultaneous determination of the injected
powerlaw slope, would then allow to estimate the real cutoff
energy of the injected proton/neutrons.

In Fig.~4 we show the 95\% confidence level for the expected flux
at an AGASA-like experiment at Earth, using 1000 random
realizations of GRBs in the inner Galaxy. The horizontal line
shows roughly the average value actually observed for the excess
signal by AGASA. The density distributions are from
\cite{Lepine03}, and the magnetic field distributions from
\cite{Stanev95,Beck96,HaMoRo99,Beck00,GMTParis01}.  In these
specific simulations the cross-sections were taken from
\cite{PDG02}.

The spatial distribution of the neutrons should mimic roughly a
folding of the decay distance at some energy of the presumed
original neutron with the matter distribution. Thus it should be
similar to the gamma ray emission and the FIR emission along the
Galactic plane.  We note that the estimated residence time at
$10^{18}$ eV energies corresponds to a real travel distance via
diffusion of only about 2 to 3 kpc, while the propagation and
interaction distance (i.e. the length along the meandering path)
is 30 kpc.

It is interesting to consider the time evolution of such a neutron
flux:
  During a first phase protons resulting from neutron decay interact to
produce new neutrons, and so steadily increase the observable
neutron flux.  Second, we begin to use up the reservoir of
protons, so the flux of neutrons decreases slowly, with
$t^{-3/2}$.  And third, the flux begins to decay with $t^{-5/2}$,
as protons leak out from the probable interaction volume.

The spatial distribution strongly depends on the specific
orientation and the exact location of the original double jet of
the GRB - if that is the best model.   Of course, it is very
unlikely, that the GRB jet would point exactly at Earth - that
might be damaging to us, since this is the same power in a few
seconds as the Sun gives in minutes, and in a less benign form.
Therefore, if we see several GRB events and their consequences
superimposed on each other, then the discrepancy between AGASA and
SUGAR in spatial distribution might be partially explainable.

\section{Conclusions}

We have shown that it is plausible that the observed AGASA excess
of events near $10^{18}$ eV energies coming to us from the GC
region is due to the last few GRB events in the Galaxy.  We
predict a corresponding flux in photons and neutrinos.

In fact, if the predicted details can be confirmed, we will have
established i) that GRB cosmic ray signature can be detected, ii)
the cosmic ray production of GRBs to be of order $10^{51}$ erg,
iii) that their particle energy extends to at least $6 \times
10^{18}$ eV, iv)  that the maximum particle energy can be
estimated with a measurement of both neutrons and photons, as well
as the slope of the injection spectrum, and v)  that their
contribution to the overall energetics of Galactic cosmic rays is
minor.  To check this will be a major contribution of the Pierre
Auger Observatory whose southern part is ideally located to
observe the GC region \cite{AUGER}.  The combination of
fluorescence and surface detectors of this experiment allow
measurements in the energy region from several $10^{17}$ eV to the
highest energies.

\section{Acknowledgements}

The authors appreciate comments from A. Watson  and T. Kellmann.
PLB is mainly supported through the AUGER theory and membership
grant 05 CU1ERA/3 through DESY/BMBF (Germany); GMT is also
supported by FAPESP and CNPq.

\end{document}